# Feedback Systems for Linear Colliders

L. Hendrickson *et al.*

Invited talk presented at IEEE Particle Accelerator Conference (PAC99),
3/29/99ã4/2/99, New York, NY, USA

*Stanford Linear Accelerator Center, Stanford University, Stanford, CA 94309*

Work supported by Department of Energy contract DE AC03 76SF00515.

# FEEDBACK SYSTEMS FOR LINEAR COLLIDERS[*]

L. Hendrickson#, P. Grossberg, T. Himel, M. Minty, N. Phinney, P. Raimondi,
T. Raubenheimer, H. Shoaee, P. Tenenbaum,  SLAC, Stanford, CA

*Abstract*

Feedback systems are essential for stable operation of a linear collider, providing a cost-effective method for relaxing tight tolerances. In the Stanford Linear Collider (SLC), feedback controls beam parameters such as trajectory, energy, and intensity throughout the accelerator. A novel dithering optimization system which adjusts final focus parameters to maximize luminosity contributed to achieving record performance in the 1997-98 run. Performance limitations of the steering feedback have been investigated, and improvements have been made.

For the Next Linear Collider (NLC), extensive feedback systems are planned as an integral part of the design. Feedback requirements for JLC (the Japanese Linear Collider) are essentially identical to NLC; some of the TESLA requirements are similar but there are significant differences. For NLC, algorithms which incorporate improvements upon the SLC implementation are being prototyped. Specialized systems for the damping rings, RF and interaction point will operate at high bandwidth and fast response. To correct for the motion of individual bunches within a train, both feedforward and feedback systems are planned. SLC experience has shown that feedback systems are an invaluable operational tool for decoupling systems, allowing precision tuning, and providing pulse-to-pulse diagnostics. Feedback systems for the NLC will incorporate the key SLC features and the benefits of advancing technologies.

## 1 OPERATIONAL ISSUES

Linear colliders have severe operational challenges, and feedback systems are an essential tool. Feedback systems distributed throughout the machine allow less-invasive tuning procedures, so that when upstream parameters are modified, downstream feedback stabilizes the beam, allowing routine operation to continue. Feedback is also invaluable in facilitating quick startup after outages. A robust design ensures that the feedback is not confused when the beam returns after an outage, and that control devices are not moved in the absence of beam. Feedback is a useful element in higher-order tuning applications. In the SLC linac, the emittance tuning packages move feedback setpoints to create an oscillation in the linac which is closed by a downstream feedback. In the final focus, the built-in averaging capabilities of the optimization feedback enable it to be used as a high-resolution measurement device for complex tuning procedures. A variety of diagnostics are provided for beam jitter, fit quality and other information, and the feedback system is also used to take a snapshot of data for diagnosing machine trips.

## 2 SLC FEEDBACK

The SLC feedback system provides a valuable base of experience for future systems. A generalized, database driven beam based system [1] was implemented starting in 1990. The system was expanded to control a variety of beam parameters in every major region of the machine. At the end of the 1998 run, there were 50 control loops for the SLC alone, and another 25 to support other programs such as the PEPII B factory, Final Focus Test Beam and fixed target experiments. In general the data is available to the feedback at the pulsed rate of 120 Hz, although many loops run more slowly due to CPU limitations and other considerations.

### 2.1 Basic Feedback Systems

In the injector, feedback controls a variety of intensity-related parameters associated with the polarized electron gun, including laser voltages and kicker timing. A higher order system controls the asymmetry between the intensity of left and right polarized beam averaging several thousand pulses.

Steering feedback systems are found throughout the machine, reading beam position monitors (BPMs) and moving correctors. Several performance limitations with these systems are discussed in a later section. A cascade system was developed to minimize overcorrection associated with multiple feedback loops in a beamline which all control the same parameters [2].

The energy of the SLC is controlled in a variety of locations, typically by reading BPMs in a high-dispersion region and moving phase shifters or controlling klystron amplitudes. A hardware based feedforward system coordinates with the beam based feedback to compensate for the energy change due to intensity fluctuations. The intensity is measured while the beam is in the damping ring and communicated to the energy feedback system in the linac for correction on the same pulse.

At the interaction point, a specialized feedback keeps the beams in collision using the beam-beam deflection measured from the BPMs. The beam-beam separation is determined by normalizing the deflection to compensate

[*]Work supported by the U.S. Dept. of Energy under contract DE-AC03-76SF00515

# Email: ljh@slac.stanford.edu

for intensity fluctuations, and fast pulsed correctors allow full 120 Hz control.

## 2.2 Feedback Architecture and Calculations

The SLC feedback system is generalized and database driven so that feedback loops can often be added without additional software. The feedback is designed to use the components (CPU, BPMs, correctors, etc.) of the existing control system and dedicated hardware is not required except where speed is essential. A fast point-to-point network has been developed to communicate between different microprocessors, supporting feedback loops with elements from more than one region. The feedback calculations use matrices which are generated offline and stored in an online database. The matrix design is based on the state space formalism of digital control theory, using Kalman filters and Linear Quadratic Gaussian regulators [3]. The noise model includes a combination of white noise and low frequency noise, with the goal of achieving a robust system with good step response.

## 2.3 Specialized Frequency Control

The SLC is a 120 Hz pulsed machine, with a slight difference in even and odd pulses (timeslots) due to the AC power sources, causing 60 Hz energy or trajectory oscillations. A feedback system stabilizes the energy difference between the two timeslots using specialized logic that calculates and controls the average and difference of odd and even pulses, resulting in excellent damping at and near the Nyquist frequency.

A 59 Hz oscillation in the beam position was caused by pump vibrations in the linac, and the downsampled 20 Hz loops resulted in aliasing to 1 Hz. The rate of some linac feedback systems was modified to minimize this aliasing. To damp the 59 Hz, feedback with the timeslot control algorithm was implemented at the end of the linac and at the interaction point. Performance tests showed that the 59 Hz oscillation was damped by a factor of 5.

## 2.4 Optimization Feedback

Optimization feedback has been implemented for the SLC which uses automated subtolerance excitation techniques in order to determine an optimal setting for a nonlinear system. The first optimization feedback controlled energy spread by varying phase offsets [4]. The phase offset was moved up and down by a small amount, averaging over many pulses and then the slope of a parabola was calculated to determine the optimal setting. This application was found useful at first but the minimum control bit size available in the hardware was too invasive for routine operation, and this automated procedure was eventually discontinued. A similar system was developed to minimize the linac jitter by optimizing the damping ring extraction kicker timing. This system worked well in principle, but was used intermittently and never remained in routine operation.

In the final focus of the SLC, five orthogonal parameters for each of the two beams are routinely tuned to maximize the luminosity. Originally, a scan method was employed, where a linear combination of devices was moved through a series of settings, the beam size was measured at each point, and a parabola was fitted to the square of the beam size to determine the optimal setting. The beam size was measured using beam-beam deflection scans. With smaller beam sizes, the measurement resolution became a significant issue [5], leading to an estimated 20-40% luminosity loss due to mistuning. Furthermore, the tuning procedures required a significant amount of operator time and attention, and the results were dependent upon individual judgement, resulting in variable luminosity. At higher luminosity, monitors became available to return a statistically significant signal which is proportional to the luminosity on a pulse to pulse basis. These monitors use beamstrahlung radiation from the beam-beam interaction and also wide angle radiated Bhabha scattering events.

A fully automated feedback system [6] was developed to dither each linear combination of devices through three settings, averaging the luminosity signal over many pulses. Figure 1 shows the dithering process in operation; as the dither knob is moved, the luminosity monitors show a response. Under most conditions, the Bhabha monitor could have excessive luminosity-related backgrounds, and the beamstrahlung detector provided better results. The offset of a parabola was calculated to determine the optimal point. A scheduling process determined which of the ten parameters was to be optimized at a given time, based upon user-selected scheduling and tolerances. The dithering process required only a few percent luminosity degradation, while improving the resolution by a factor of 5 to 10. Overall this resulted in a significant net luminosity increase, while allowing the operations staff to concentrate on other tuning.

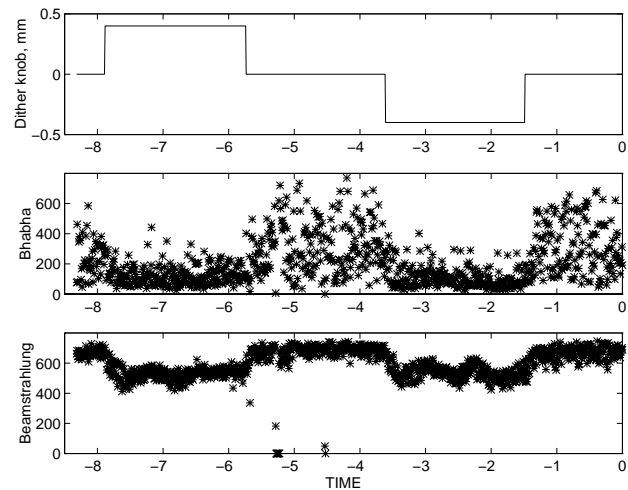

Fig 1: Dithering Process for Luminosity Optimization Feedback

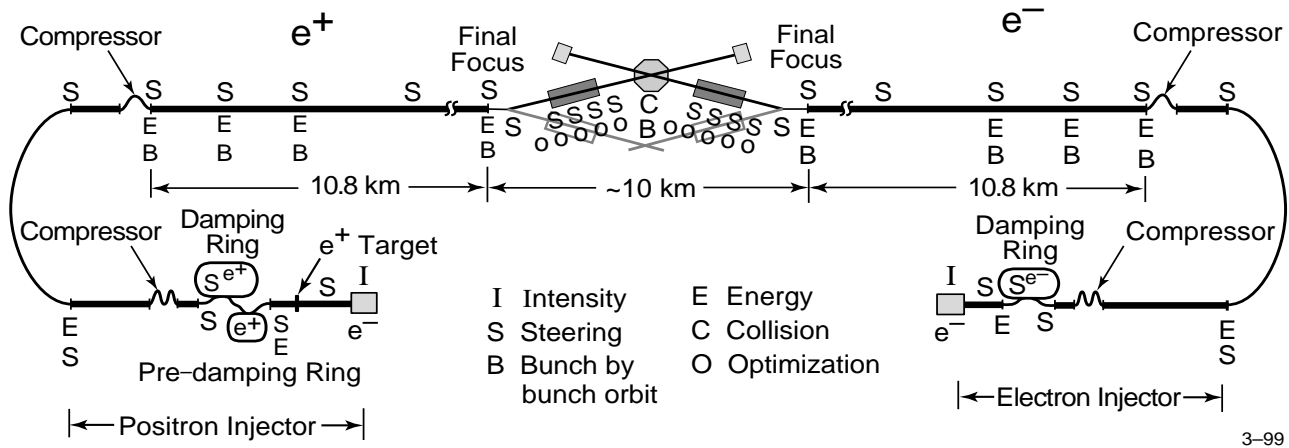

Fig 2: NLC Schematic with Feedback Loops Shown

## 3 NLC FEEDBACK

The NLC requires several new types of feedback systems in addition to the capabilities developed for the SLC. The control system must provide the CPU and network bandwidth to support flexible feedback communications at the full machine rate. Performance improvements are also needed to remedy problems found at the SLC. Figure 2 shows locations where feedback systems are planned.

### 3.1 Bunch-to-bunch Feedback

The NLC will have a pulse train of 95 bunches. In addition to controlling beam parameters for the mean of the bunches in the train, feedback systems are required to correct the differences of intensity, energy and position along the train. Because the interbunch spacing is only 2.8 nanoseconds, most of these loops are not designed to respond in an interbunch period, but would typically run at the pulse rate of 120 Hz, correcting the shape of the train on a later pulse. Specialized beam position monitors must return information on each bunch of the train, and pulsed actuators are required to have enough bandwidth to adjust the individual bunches.

An ultrafast feedback is being studied to bring the beams into collision for subsequent bunches of the pulse train [7]. Separate dedicated hardware based systems would control the horizontal and vertical planes. The relative beam offset would be measured by an outgoing beam position monitor near the interaction point, and the control actuator would be a weak, fast kicker controlling the incoming position of the other beam. This configuration minimizes delays for signal travel as both measurement and control are in the same location. A pilot bunch which is sent through undeflected, without the opposing beam, would provide a reference for deflection calculations. The measured deflection of the first colliding bunches can then be used to calculate a correction for the remainder of the train. The gain factor to convert deflection angle to position would be set by an interface with a slower software based feedback. Preliminary estimates indicate that the feedback may be able to produce a control response within about 10 ns.

### 3.2 Damping Rings

For the NLC damping rings, longitudinal and transverse feedback systems will take advantage of architecture and algorithms from storage rings such as PEPII, APS, PSI, etc. Some of these will be dedicated high-bandwidth systems, but beam based global or local orbit feedback systems are also anticipated.

### 3.3 Additional Capabilities

Several other unusual feedback systems are planned for the NLC. To stabilize vibration of the final quadrupoles near the IP, a hybrid system may be used to span both low and high frequencies. An accelerometer based system measuring vibration data at 5 KHz may provide high frequency response while a beam based feedback running at 120 Hz covers low frequencies. Another feedback system must synchronize the phases of the two final focus crab cavities. Additional systems will stabilize the relative phase of the beam with respect to the RF, and control the effective fiber length for the timing system. Finally, a variety of RF-related feedback systems are planned.

### 3.4 Control System Issues

The NLC feedback system will be well integrated with the rest of the control system. Feedback will share measurements with users, so that requested measurements do not interrupt the feedback system. Sufficient CPU and networking are planned, so that feedback can generally run at the full beam rate and latency is minimized. The standard timing budget will result in a control response within two 120 Hz interpulse periods after a perturbation is seen. One pulse is allowed for digitization of the BPMs, network traffic and feedback computations. A second pulse is required to ramp power supplies and affect the beam. New BPMs are designed for high resolution, good stability and wide dynamic range.

Coordination with the main control system is essential to maximize effectiveness. Diagnostics should include fast time plots, where users can view beam parameters on a pulse to pulse basis, as well as longer-term history plots, RMS calculations and FFTs. Feedback data should be accessible to and integrated with other applications. Applications should be able to automatically turn on and off feedback loops, change setpoints and obtain control values. A variety of measurement devices and actuator types should be supported in order to allow for easy expansion. Other applications must recognise when a device is under feedback control to avoid contention. Finally, flexible networking systems must accommodate unplanned future extensions.

## 4 PERFORMANCE ISSUES

Performance limitations have been a significant area of concern with the SLC feedback system, and one of the goals for the NLC is to insure that these issues are adequately understood and addressed. Experiments using the SLC have been able to identify and characterize a variety of problems.

### 4.1 Cascade

In both the SLC and NLC, the linac is a long, straight beamline with several orbit feedback systems in sequence. Overcompensation and ringing will occur if multiple systems respond to the same disturbance. For the SLC linac, a cascade system communicates processed beam information from each feedback to the next downstream loop. The transport matrices between loops are calculated adaptively to eliminate sensitivity to optics drifts. The system was initially successful, allowing feedback to run at high gain factors with good system response. However, this algorithm assumes that the beam transport is independent of the source of a perturbation.

With higher intensity operation, wakefields and chromatic effects make the beam transport nonlinear and oscillations propagate differently depending on their point of origin. When a bunch passes off-axis through the accelerating structures, the asymmetric fields induced by the head of the bunch kick the later particles, producing a tilted distribution which remains after the centroid is corrected. In order to correctly model the beam transport, each feedback must receive information from all of the upstream loops to identify the source of the disturbance and avoid overcorrection. An improved algorithm has been successfully tested in simulation.

### 4.2 Simulation Environment

In order to evaluate feedback designs and analyze performance issues, a Matlab simulation environment was created [8]. The LIAR beam transport code [9] is called as a subroutine, realistically simulating wakefields, ground motion effects such as ATL, and other errors including alignment and field strength. The state space matrix design program was converted from MatrixX to Matlab to allow for easier simulations and evaluation of alternative algorithms.

### 4.3 Configuration for Wakefield Effects

The configuration of feedback for the NLC linac was modelled using the Matlab simulation environment. The original proposal was based on a typical SLC setup, where linac feedback loops use a minimal number of correctors immediately followed by 8-10 BPMs. Due to wakefield effects, this configuration resulted in a large orbit RMS and significant emittance growth. Figure 3 shows the vertical beam position along the linac after a perturbation has been introduced at the beginning of the linac and the feedback has converged. Each feedback perfectly flattens the orbit at the feedback BPMs, but large orbit excursions are seen between the feedback locations.

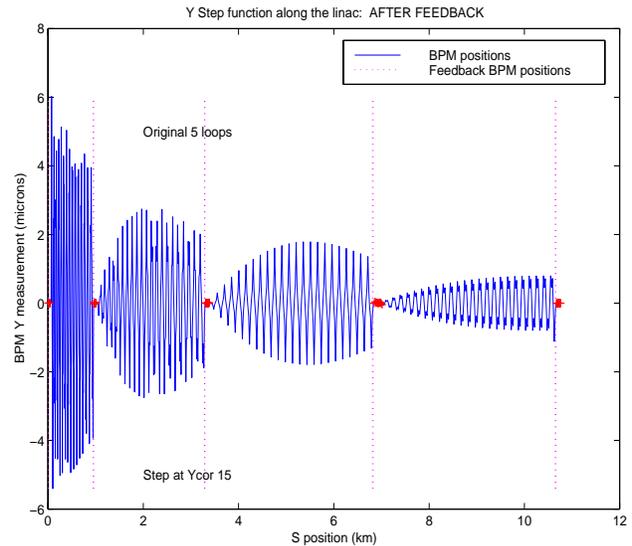

Fig 3: Linac Orbit with Original Feedback Configuration

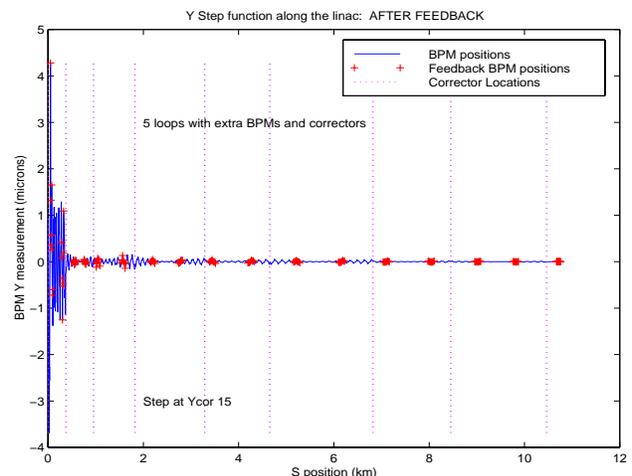

Fig 4: Linac Orbit with Improved Feedback Configuration

Several alternative configurations were evaluated with respect to orbit RMS, emittance growth, bunch shape at the end of the linac, and ATL ground motion response. Better performance was achieved by adding more beam position monitors and correctors than were originally proposed. When these additional devices are spread over long distances in a single feedback, the wakefield compensation can be further improved. However, there are practical limitations to the span of a single feedback due to imperfections in the model. For a global feedback controlling the entire linac, these errors can be significant and degrade performance. More work is needed to evaluate beam transport variation with klystron phase offsets and other errors. In the presently preferred configuration, the linac is divided among five feedback systems, each including four sets of BPMs and two sets of correctors. Figure 4 shows the resulting beam trajectory for this system, after it has responded to the same disturbance shown in figure 3.

### 4.4 Other Performance Issues

Measurements of SLC corrector response determined that the speeds of many devices were much slower than in the feedback design model. The model assumes a delay of two 120 Hz pulses for the actuators to respond but in many locations, the measured response is a ramp of 9 or more pulses. Simulations indicate this would have only a moderate effect on feedback performance if it were the only flaw. However, when combined with other imperfections, the slow correctors can have a significant performance impact.

In some areas of the SLC, the beam transport is poorly modeled. Sometimes this can be fixed by identifying and correcting accelerator errors such as klystron mis-phasing. Otherwise a calibration procedure is used to measure the beam transport and incorporate it into the feedback design.

Hardware problems such as broken correctors and erratic BPMs also degrade the feedback performance if they are not detected. Due to limited CPU and networking and a suboptimal cascade scheme, many feedback loops run slower than the full 120 Hz pulse rate, and most run with decreased feedback gain factors. This results in a system response which amplifies beam noise around 1 Hz. At the SLC, the large number of feedback loops, the many sources of imperfections, and inadequate diagnostic tools result in degraded feedback performance.

For NLC, plans are underway to ensure that these issues are adequately addressed. Additional SLC beam testing and simulations are planned. More work is needed to analyze calibration or adaption schemes for the beam transport model. In order to diagnose remaining problems a feedback performance watchdog is needed. An excellent algorithm for this could come from model-independent analysis techniques [10], which are capable of identifying the locations of broken BPMs, correctors and misbehaving feedback loops.

## 5 TESLA FEEDBACK

The TESLA Linear Collider has some significant differences from NLC. The linac uses superconducting RF where the wakefield effects are minimal, simplifying the feedback requirements. The interbunch spacing is 377 ns, compared with 2.8 ns in the NLC. This allows for orbit feedback systems to more easily correct on subsequent bunches of the train. Plans for these systems are described more fully in reference [11].

## 6 CONCLUSIONS

Linear colliders present significant challenges for controls, due to tight tolerances and stability requirements, and complex tuning procedures. Extensive feedback systems are an essential tool for operation. Much work remains to be done for future colliders to insure that the benefits of these feedback systems are fully realized, while addressing the performance issues associated with a large number of systems.

## 7 ACKNOWLEDGEMENTS

The authors wish to thank M. Breidenbach, F. J. Decker, J. Frisch, I. Reyzl, and M. C. Ross for their valuable contributions to this paper and the work it describes.

## 8 REFERENCES

[1] L. Hendrickson, et al., "Generalized Fast Feedback System in the SLC," ICALEPCS, Tsukuba, Japan, SLAC-PUB-5683 (1991).
[2] T. Himel, et al., "Adaptive Cascaded Beam-Based Feedback at the SLC," PAC, Washington, D.C., SLAC-PUB-6125 (1993).
[3] T. Himel, et al., "Use of Digital Control Theory State Space Formalism for Feedback at SLC," PAC, San Francisco, CA, SLAC-PUB-5470 (1991).
[4] M. Ross, et al., "Precise System Stabilization at SLC Using Dither Techniques," PAC, Washington, D.C., SLAC-PUB-6102 (1993).
[5] P. Emma, et al., "Limitations of Interaction-Point Spot-Size Tuning at the SLC," PAC, Vancouver, Canada, SLAC-PUB-7509 (1997).
[6] L. Hendrickson,et al., "Luminosity Optimization Feedback in the SLC," ICALEPCS,Beijing,China, SLAC-PUB-8027 (1999).
[7] M. Breidenbach, private communication (1999).
[8] P. Tenenbaum, et al., "Use of Simulation Programs for the Modelling of the Next Linear Collider," Proceedings of this conference (1999).
[9] R. Assmann, et al., "The Computer program LIAR for the simulation and modeling of high performance linacs," PAC, Vancouver, Canada (1997).
[10] J. Irwin, et al., "Model-Independent Beam Dynamics Analysis," Proceedings of this conference (1999).
[11] I. Reyzl, et al., "Fast Feedback Systems for Orbit Correction in the TESLA Linear Collider," Proceedings of this conference (1999).